%% file: main.tex
\documentclass[useAMS,usenatbib]{mn2e}
\usepackage[colorlinks,linkcolor=blue,citecolor=blue]{hyperref}

\usepackage{amssymb}
\usepackage{amsmath}
\usepackage{txfonts}
\usepackage{graphicx}
\usepackage{subcaption} 
\usepackage{epsfig}
\usepackage{xspace}
\usepackage{url}
\usepackage[colorinlistoftodos]{todonotes}

\usepackage{natbib}
\def\xmm{\textit{XMM-Newton}\xspace}

\def\flat{\textit{Fermi/LAT}\xspace}

\def\psrb{PSR~B1259-63\xspace}

\def\halpha{H\,$\alpha$\xspace}

\title[2021 periastron passage of  PSR B1259-63 ]{The radio to GeV picture of  PSR B1259-63 during the 2021 periastron passage.}
\author[Chernyakova et al]{M.~Chernyakova$^{1,2}$, D.~Malyshev$^3$, B. van Soelen$^{4}$, S. Mc Keague 
 $^{1}$, S.~P.~O'Sullivan $^{5}$,
 \newauthor{and D. Buckley $^{4,6,7}$}\\
 $^{1}$  School of Physical Sciences and Centre for Astrophysics \& Relativity, Dublin City University\\
$^{2}$  Dublin Institute for Advanced Studies, 31 Fitzwilliam Place, D02 XF86 Dublin 2,  Ireland \\
$^{3}$  Institut f{\"u}r Astronomie und Astrophysik T{\"u}bingen, Universit{\"a}t T{\"u}bingen, Sand 1, \mbox{D-72076 T{\"u}bingen, Germany}\\
$^{4}$  Department of Physics, University of the Free State, P.O. 
 Box 339, Bloemfontein 9300, South Africa\\
 $^{5}$ Departamento de Física de la Tierra y Astrofísica \& IPARCOS-UCM, Universidad Complutense de Madrid, 28040 Madrid, Spain\\
$^{6}$ South African Astronomical Observatory \& Southern African Large Telescope, PO Box 9, Observatory 7935, Cape Town, South Africa\\
$^{7}$Department of Astronomy, University of Cape Town, Private Bag X3, Rondebosch 7701, South Africa\\}
 
\begin{document}

\maketitle
\begin{abstract}
\psrb is a gamma-ray binary system with a radio pulsar orbiting an O9.5Ve star, LS 2883, with a period of $\sim3.4$~yr. Close to the periastron the system is detected at all wavelengths, from radio to the TeV energies. The emission in this time period is believed to originate from the interaction of LS~2883 and pulsar's outflows. The observations of 4 periastra passages taken in 2010-2021 show strong correlation of the radio and X-ray light curves with two peaks just before and after the periastron. The observations of the latest 2021 periastron passage reveal the presence of the 3rd X-ray peak and subsequent disappearance of radio/X-ray flux correlation. 
In this paper we present the results of our optical, radio and X-ray observational campaigns on \psrb performed in 2021 accompanied with the analysis of the publicly available GeV \flat data. We compare the properties of different periastron passages, discuss the obtained results and show that they can be explained in terms of the 2-zone emission 
model proposed by us previously.
\end{abstract}

\begin{keywords}
gamma rays: general; pulsars: individual: PSR B1259-63; stars: emission-line, Be; X-rays:
binaries; X-rays: individual: PSR B1259-63; radiation mechanisms: non-thermal
\end{keywords}

\section{Introduction}
 \psrb is a gamma-ray binary with a 47.76~ms radio pulsar orbiting a O9.5Ve  star (LS~2883)  with a period of $\sim 1236.7$~days  in a highly eccentric orbit ($e\sim 0.87$) \citep{1992ApJ...387L..37J,Negueruela-2001,shannon14}.  Based on the parallax data in the Gaia DR2 Archive~\citep{Gaia2018} the distance to the system is $2.39 \pm 0.19$~kpc, which is consistent with the value of $2.6^{+0.4}_{-0.3}$ kpc  reported by \citet{PSRB1259-2018_distance}.

 \psrb was discovered in 1992 during the Parkes Galactic plane survey \citep{1992ApJ...387L..37J,1992MNRAS.255..401J}. Further observations revealed that the pulsed radio emission is completely absorbed from around 20 days before to 20 days after periastron, being obscured by the circumstellar disc of the Be star which is inclined to the orbital plane~\citep{Wang2004}.\footnote{While LS\,2883 is an O-type star, the Oe and Be stars are generally grouped together and referred to as Be stars.}

The interaction of the pulsar wind with the equatorial wind of the Be star leads to the appearance of unpulsed radio emission about 20 days before the periastron, quickly rising to a flux exceeding the pulsed flux by a factor of several tens, and lasts for at least 100 hundred days after periastron. The light curve of this unpulsed emission  shows two peaks associated with the time of pulsar crossing the plane of the circumstellar disk, before and after the periastron  \citep{2002MNRAS.336.1201C,2005MNRAS.358.1069J}. A similar two-peak structure is also observed in X-rays \citep[see e.g.][]{Chernyakova2006,Chernyakova2009}.

During the periastron passage, \psrb is also detected at high and very high energies \citep{Chernyakova21_psrb, hess_psrb2020}.
  {Current H.E.S.S. observations indicate that the TeV light curve may also have a two peak structure, similar to what is observed at radio and X-ray energies \citep{hess_psrb2020}, but more sensitive observations are needed to confirm this.}
Hopefully CTA will address this issue in the very near future \citep{CTArev2019}. The GeV emission, however, shows a very different behaviour and is characterised by a strong flare   { with an average luminosity approaching that of the pulsar spin-down luminosity}, occurring at least 10 days after the second X-ray peak with a position varying from periastron to periastron with no obvious counterpart at other wavelength \citep[e.g.][]{Chernyakova21_psrb}.

  {Several models have been proposed so far to explain the observed multiwavelength emission from this system.
The broadband emission has been proposed to originate from electrons accelerated in the shock that forms between the pulsar wind and stellar wind and is produced by synchrotron and IC radiation, e.g. \citet{ Dubus06, bogovalov08, khan11,2014MNRAS.439..432C,Chen2019}. }

  { In \citet{chernyakova15}  the GeV flare was explained as a result of synchrotron cooling of monoenergetic relativistic electrons injected into the system during the disruption of the stellar disc.
The comptonization of unshocked pulsar wind particles~\citep{khan12}  as well as Doppler boosting~\citep{dubus_cerutti10, kong12} were also suggested to play a role in producing the GeV flares. \citet{2017ApJ...844..114Y}  proposed that the GeV flare can be a result of the transition between the ejector, propeller, and accretor phases. In this model the compact object is  working as an ejector all along its orbit and being powered by the propeller effect when it is close to  periastron, in a so-called flip-flop state. 
}

  { Despite the variety of the models mentioned above, none of them can explain or predict the ultra-luminous short-time subflares discovered during 2017 periastron passage.} It turned out that the GeV flare consists of a large number of bright sub-flares which were as short as 15 minutes and had luminosities exceeding the pulsar spin-down luminosity by more than an order of magnitude \citep{2018ApJ...863...27J,Tam2018,Ghang2018, Chernyakova20_psrb, Chernyakova21_psrb}. 
To explain these observations \citet{Chernyakova20_psrb} propose a model in which the observed X-ray and TeV emission is generated by the highly relativistic electrons of the pulsar wind strongly accelerated at the apex of the shock formed between the pulsar and stellar winds.
The GeV emission, in turn, is produced by the IC emission of weakly accelerated electrons flowing from the system along the shock, with a possible addition of bremsstrahlung emission on the clumps of the stellar wind material which penetrated beyond the shock cone. In this case one can expect to see a magnification of the GeV emission when the cone is looking in the direction of the observer.

To test this model we have organised an intensive multi-wavelength campaign to observe the 2021 periastron passage \citep{Chernyakova21_psrb}.  This campaign revealed a number of unexpected features in the system, like a significant delay of the GeV flare, the presence of a third X-ray peak and disappearance of the correlation of the radio and X-ray data observed during the second X-ray peak.

In the current paper we present more details on the radio behaviour of the system including data at 9 GHz in addition to the 5.5 GHz presented earlier, the evolution of the pulsed flux and the polarisation properties (Section 2.1). In Section 2.2 we present 2021 SALT optical  observations and compare them with the reanalysed 2014 and 2017 data.  In Section 2.3 we present the analysis of newly available \xmm data. In addition to that we have reanalysed all available \flat data in order to compare the main properties of the GeV flare (mainly spectrum and timing structure) from different periastron passages (Section 2.4). Broad band spectral modeling of the new data is presented in Section 3, and Section 4 is devoted to the discussion of the relation between radio and X-ray components in the source. Finally the discussion and conclusions are presented in Section 5.


\section{Data Analysis}
\subsection{Radio Data}
\begin{figure*}
\includegraphics[width=0.32\linewidth]{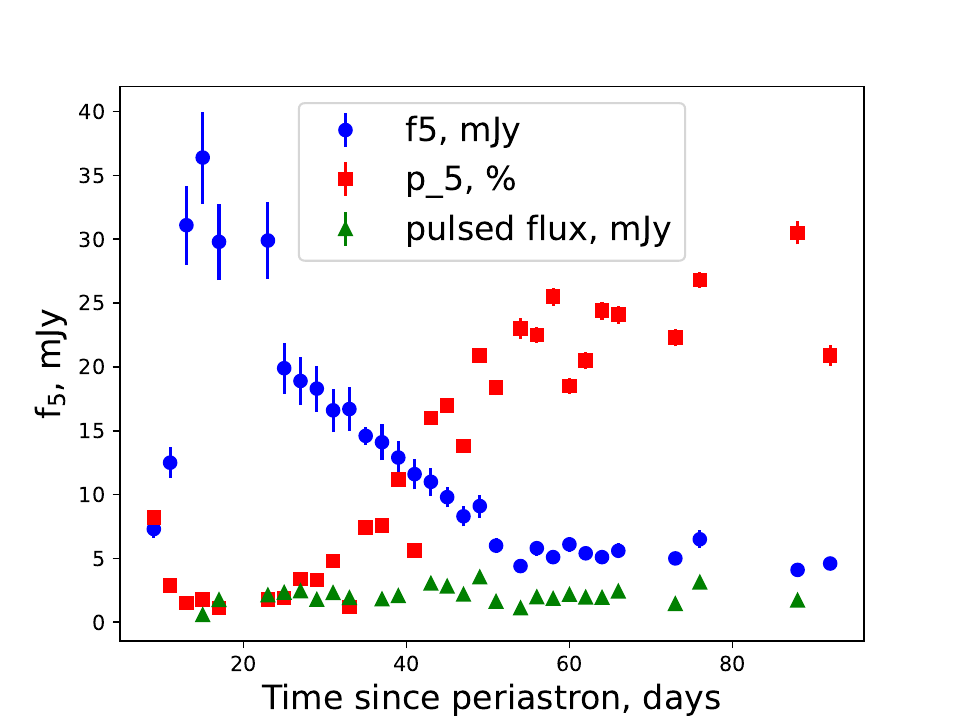}
\includegraphics[width=0.32\linewidth]{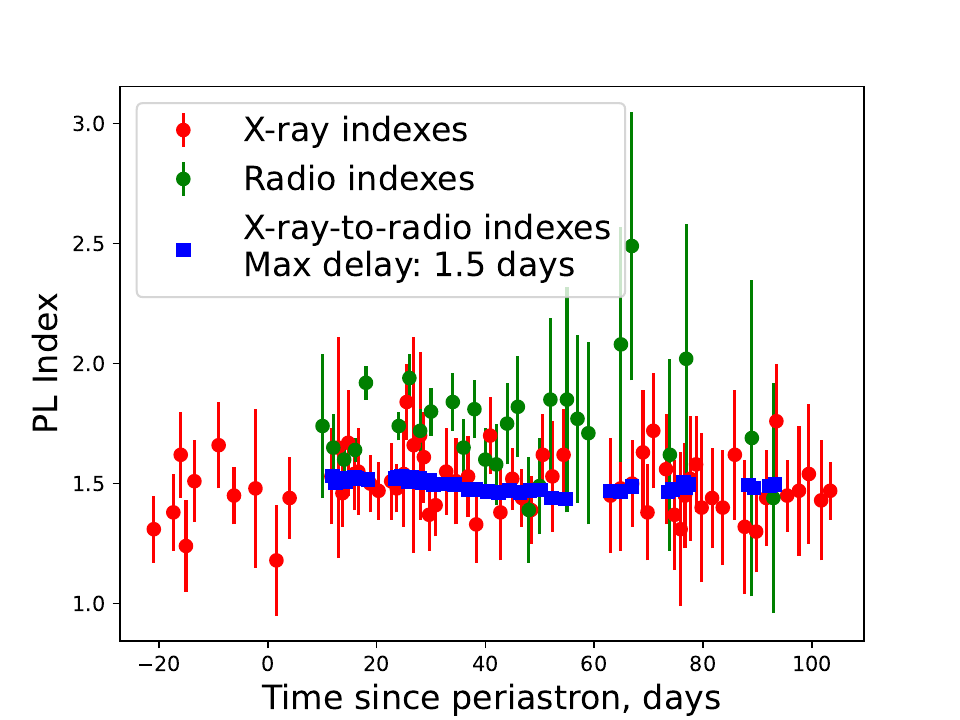}
\includegraphics[width=0.32\linewidth]{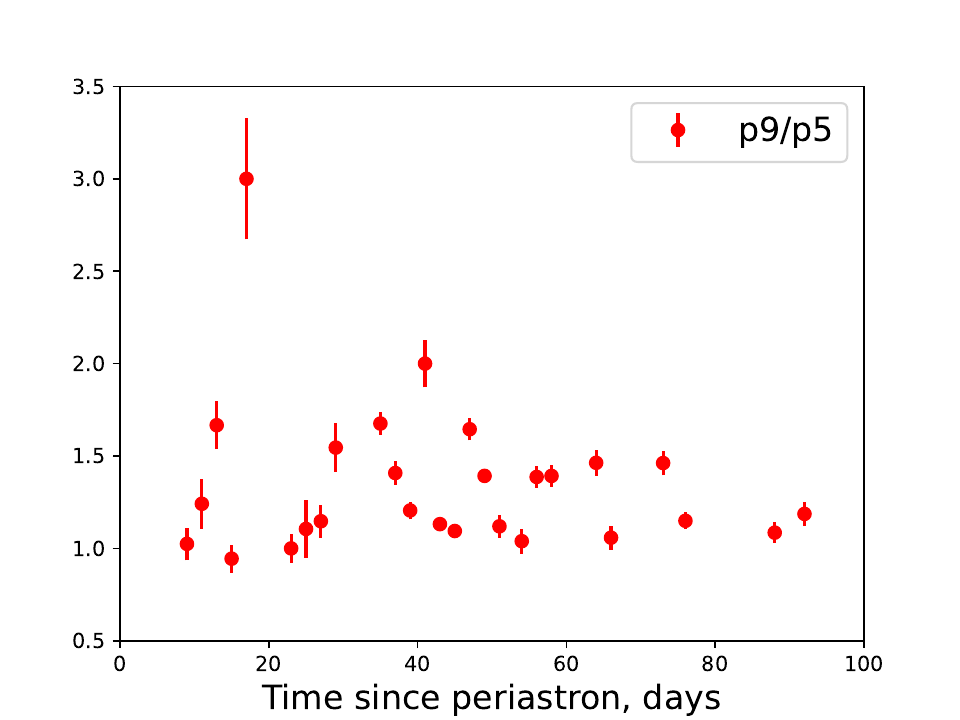}
\caption{
\textit{Left:} Evolution of the total flux, pulsed flux and degree  of polarization at 5.5 GHz.
\textit{Middle:} Evolution of X-ray (red) and Radio (green, $1- \alpha$) photon indexes around the periastron. In blue we show photon indexes that would explain the overall radio to X-ray spectrum   {(using data points separated by no more than 1.5~days)}.
\textit{Right:} Ratio of degrees  of polarization at 9 GHz and 5.5 GHz. One outlier happened 34 days after the periastron is not shown at a value of 10.7.
}
\label{fig:radioflux}
\end{figure*}

\subsubsection{Total radio emission from \psrb}
The radio data were obtained with the Australia Telescope Compact Array (ATCA) using the 4 cm observing band, starting from $\sim$10\,days after the 2021 periastron passage. For the initial calibration, we followed the procedure as described in \citet{Chernyakova21_psrb}, using the {\sc miriad} software package, with B1934$-$638 as the bandpass and flux calibrator and J1322$-$6532 as the  phase calibrator, which we also used for correcting for the instrumental polarization following standard techniques \citep{sault1995}. The {\sc miriad} task {\sc uvflux} was then used to determine the total flux density and the degree of polarization of PSRB1259$-$63, using {\sc options=uvpol} and an inner uv-cut of 20 kilo-wavelengths. The inner uv-cut was employed to minimise contamination of the flux from other sources in the field of view and any diffuse Galactic emission. This provided measurements at both 5.5 GHz and 9 GHz (using the full 2 GHz of bandwidth in both cases), as shown in Table \ref{tab:radiodata} and Fig. \ref{fig:radioflux} (left panel). The uncertainties on individual values were obtain from adding in quadrature the rms noise of the visibilties and the nominal absolute flux density uncertainty of 10\%. Some total emission values at 9 GHz are missing due to the failure of the calibration on those days (59286.88, 59315.77, 59317.77). In the case of the pulsar flux, it was not detectable during the first three observations, while for some proceeding days, reliable flux densities could not be extracted, either at 5.5 GHz (59290.85), or both 5.5 and 9 GHz (59296.85, 59347.78). The first detection of the pulsed emission occurred $\sim$ 16 days after the periastron. 

\begin{table*}
 \centering
    \caption{ATCA observations of \psrb in 2021. The table summarizes the modified Julian date of the observation (MJD), time since periastron (t$_p$=MJD-59254.867), flux density at 5.5 GHz (F$_5$, mJy), pulsed flux density  at 5.5 GHz (F$_{p,5}$, mJy), degree of polarisation at 5.5 GHz (p$_{5}$, in \%),   flux at 9 GHz (F$_9$, mJy), pulsed flux  at 9 GHz (F$_{p,9}$, mJy), degree of polarisation at 9 GHz (p$_{9}$, in \%) and the spectral index $\alpha$ }
    \begin{tabular}{c|c|c|c|c|c|c|c|c}
    MJD& t$_p$(d)& F$_5$& F$_{p,5}$&p$_5$& F$_9$& F$_{p,9}$&p$_9$ &$\alpha$\\
    \hline
59264.94 &  10.07 & 7.30 $\pm$  0.70 & -- &8.20 $\pm$  0.40 &  5.00 $\pm$  0.90 & -- & 8.40 $\pm$  0.60                                                   &  $-0.74 \pm 0.3$ \\
59266.92 &  12.05 & 12.50 $\pm$  1.20 & -- &2.90 $\pm$  0.20 &  9.00 $\pm$  1.10 & -- & 3.60 $\pm$  0.30                                                  &  $-0.65 \pm 0.14$ \\
59268.92 &  14.05 & 31.10 $\pm$  3.10 & -- &1.50 $\pm$  0.10 &  23.20 $\pm$  2.40 & -- & 2.50 $\pm$  0.10                                                 &  $-0.60 \pm 0.05$ \\
59270.92 &  16.05 & 36.40 $\pm$  3.60 & 0.58 $\pm$  0.06 &1.80 $\pm$  0.10 &  26.60 $\pm$  2.80 & 0.30 $\pm$  0.03 & 1.70 $\pm$  0.10                     &  $-0.64 \pm 0.05$ \\
59272.92 &  18.05 & 29.80 $\pm$  3.00 & 1.75 $\pm$  0.17 &1.10 $\pm$  0.10 &  19.00 $\pm$  2.10 & 0.49 $\pm$  0.05 & 3.30 $\pm$  0.20                     &  $-0.92 \pm 0.07$ \\
59278.94 &  24.07 & 29.90 $\pm$  3.00 & 2.12 $\pm$  0.21 &1.80 $\pm$  0.10 &  20.80 $\pm$  2.20 & 0.70 $\pm$  0.07 & 1.80 $\pm$  0.10                     &  $-0.74 \pm 0.06$ \\
59280.90 &  26.03 & 19.90 $\pm$  2.00 & 2.32 $\pm$  0.23 &1.90 $\pm$  0.20 &  12.50 $\pm$  1.70 & 0.46 $\pm$  0.05 & 2.10 $\pm$  0.20                     &  $-0.94 \pm 0.1$ \\
59282.88 &  28.01 & 18.90 $\pm$  1.90 & 2.45 $\pm$  0.25 &3.40 $\pm$  0.20 &  13.30 $\pm$  1.50 & 1.51 $\pm$  0.15 & 3.90 $\pm$  0.20                     &  $-0.72 \pm 0.09$ \\
59284.88 &  30.01 & 18.30 $\pm$  1.80 & 1.77 $\pm$  0.18 &3.30 $\pm$  0.20 &  12.40 $\pm$  1.40 & 0.79 $\pm$  0.08 & 5.10 $\pm$  0.30                     &  $-0.80 \pm 0.10$ \\
59286.88 &  32.01 & 16.60 $\pm$  1.70 & 2.30 $\pm$  0.23 &4.80 $\pm$  0.20 &  -- & 0.87 $\pm$  0.09 & --                                                  &  -- \\
59288.88 &  34.01 & 16.70 $\pm$  1.70 & 1.91 $\pm$  0.19 &1.20 $\pm$  0.20 &  11.00 $\pm$  1.30 & 0.91 $\pm$  0.09 & 12.80 $\pm$  0.30                    &  $-0.84 \pm 0.12$ \\
59290.85 &  35.99 & 14.60 $\pm$  1.70 & -- &7.40 $\pm$  0.20 &  10.60 $\pm$  1.20 & 1.28 $\pm$  0.13 & 12.40 $\pm$  0.30                                  &  $-0.65 \pm 0.12$ \\
59292.85 & 37.99 & 14.10 $\pm$  1.40 & 1.81 $\pm$  0.18 & 7.60 $\pm$  0.20 & 9.50 $\pm$  1.20&0.79 $\pm$  0.08 & 10.70 $\pm$  0.40                        &  $-0.81 \pm 0.12$ \\
59294.85 & 39.99 & 12.90 $\pm$  1.30 & 2.07 $\pm$  0.21 & 11.20 $\pm$  0.30 & 9.70 $\pm$  1.10&1.69 $\pm$  0.17 & 13.50 $\pm$  0.40                       &  $-0.60 \pm 0.13$ \\
59296.85 & 41.99 & 11.60 $\pm$  1.20 & -- & 5.60 $\pm$  0.30 & 8.70 $\pm$  1.10&-- & 11.20 $\pm$  0.40                                                    &  $-0.58 \pm 0.17$ \\
59298.85 & 43.99 & 11.00 $\pm$  1.10 & 3.04 $\pm$  0.30 & 16.00 $\pm$  0.30 & 7.60 $\pm$  1.00&1.81 $\pm$  0.18 & 18.10 $\pm$  0.50                       &  $-0.75 \pm 0.17$ \\
59300.83 & 45.97 & 9.80 $\pm$  1.00 & 2.82 $\pm$  0.28 & 17.00 $\pm$  0.40 & 6.50 $\pm$  0.90&1.74 $\pm$  0.17 & 18.60 $\pm$  0.50                        &  $-0.82 \pm 0.21$ \\
59302.83 & 47.97 & 8.30 $\pm$  0.80 & 2.18 $\pm$  0.22 & 13.80 $\pm$  0.40 & 6.80 $\pm$  1.00&2.12 $\pm$  0.21 & 22.70 $\pm$  0.50                        &  $-0.39 \pm 0.22$ \\
59304.81 & 49.95 & 9.10 $\pm$  0.90 & 3.53 $\pm$  0.35 & 20.90 $\pm$  0.40 & 7.20 $\pm$  1.00&3.33 $\pm$  0.33 & 29.10 $\pm$  0.50                        &  $-0.49 \pm 0.20$ \\
59306.81 & 51.95 & 6.00 $\pm$  0.60 & 1.62 $\pm$  0.16 & 18.40 $\pm$  0.60 & 3.90 $\pm$  0.80&0.97 $\pm$  0.10 & 20.60 $\pm$  0.90                        &  $-0.85 \pm 0.34$ \\
59309.85 & 54.99 & 4.40 $\pm$  0.40 & 1.11 $\pm$  0.11 & 23.00 $\pm$  0.80 & 2.90 $\pm$  0.70&0.42 $\pm$  0.04 & 23.90 $\pm$  1.30                        &  $-0.85 \pm 0.47$ \\
59311.79 & 56.92 & 5.80 $\pm$  0.60 & 1.97 $\pm$  0.20 & 22.50 $\pm$  0.60 & 3.90 $\pm$  0.80&1.38 $\pm$  0.14 & 31.20 $\pm$  1.00                        &  $-0.77 \pm 0.35$ \\
59313.79 & 58.92 & 5.10 $\pm$  0.50 & 1.85 $\pm$  0.18 & 25.50 $\pm$  0.70 & 3.60 $\pm$  0.80&1.32 $\pm$  0.13 & 35.50 $\pm$  1.10                        &  $-0.71 \pm 0.38$ \\
59315.77 & 60.90 & 6.10 $\pm$  0.60 & 2.17 $\pm$  0.22 & 18.50 $\pm$  0.60 & --&2.68 $\pm$  0.27 & --                                                     &  -- \\
59317.77 & 62.90 & 5.40 $\pm$  0.50 & 1.95 $\pm$  0.20 & 20.50 $\pm$  0.70 & --&2.64 $\pm$  0.26 & --                                                     &  -- \\
59319.77 & 64.90 & 5.10 $\pm$  0.50 & 1.93 $\pm$  0.19 & 24.40 $\pm$  0.70 & 3.00 $\pm$  0.30&0.93 $\pm$  0.09 & 35.70 $\pm$  1.40                        &  $-1.08 \pm 0.49$ \\
59321.77 & 66.90 & 5.60 $\pm$  0.60 & 2.43 $\pm$  0.24 & 24.10 $\pm$  0.70 & 2.70 $\pm$  0.30&0.51 $\pm$  0.05 & 25.50 $\pm$  1.40                        &  $-1.49 \pm 0.56$ \\
59328.77 & 73.90 & 5.00 $\pm$  0.50 & 1.44 $\pm$  0.14 & 22.30 $\pm$  0.70 & 3.70 $\pm$  0.70&1.69 $\pm$  0.17 & 32.60 $\pm$  1.10                        &  $-0.62 \pm 0.4$ \\  
59331.77  & 76.90& 6.50 $\pm$  0.70 & 3.13 $\pm$  0.31 &26.80 $\pm$  0.60 &  3.90 $\pm$  0.80 & 1.96 $\pm$  0.20 & 30.80 $\pm$  1.00                      &  $-1.02 \pm 0.56$ \\
59343.83  &  88.96& 4.10 $\pm$  0.40 & 1.72 $\pm$  0.17 &30.50 $\pm$  0.90 &  2.90 $\pm$  0.30 & 1.49 $\pm$  0.15 & 33.10 $\pm$  1.40                     &  $-0.69 \pm 0.66$ \\
59347.78  & 92.91& 4.60 $\pm$  0.50 & -- &20.90 $\pm$  0.80 &  3.70 $\pm$  0.80 & -- & 24.80 $\pm$  1.00                                                  &  $-0.44 \pm 0.48$ \\ \hline
    
    \end{tabular}
    \label{tab:radiodata}
\end{table*}

The radio spectral index of the total flux $\alpha$ ($EdN/dE\propto E^{\alpha}$) between 5.5 and 9 GHz (see middle panel of Fig.~\ref{fig:radioflux}) was calculated assuming a power law spectrum interpolation between the observed data points:  
\begin{equation}
    \rm \alpha=\frac{\ln(F_{5}/F_9)}{\ln(5.5/9)},
\end{equation}
where $F_{5}$ ($F_9$) is the total flux density at 5.5 GHz (9 GHz). The statistical uncertainties $\alpha_{err}$ of the derived indexes were obtained from the Gaussian uncertainties propagation: 
\begin{equation}
\rm \alpha_{err} = \frac{1}{ln\frac{9}{5.5}}\sqrt{\left(\frac{\Delta F_{5}}{F_{5}}\right)^2+\left(\frac{\Delta F_{9}}{F_{9}}\right)^2}
\end{equation}
The average value of the spectral index is $\sim-0.76$ with the characteristic uncertainty\footnote{Please note, that this uncertainty is mostly determined by the systematic 10\% average flux uncertainty added to the data; to be conservative we estimated the characteristic uncertainty as the average over the uncertainties of the individual data points} of $\pm 0.25$  across all days for which it could be reliably measured, confirming the nature of the radio emission as optically thin synchrotron emission.

For optically thin synchrotron emission, the degree of linear polarization ($p$) depends on how uniform the magnetic field distribution is across the emission region, with a maximum possible value of $\sim$72\%. From Fig. \ref{fig:radioflux} (left panel), we can see that the polarization's degree at 5.5~GHz  $p_5$  increases substantially when the pulsar emerges from behind the disk, going from $\lesssim$5\% to $\sim$25\%. It is also notable that $p_5$ at 5.5 GHz is systematically lower than $p_9$ at 9 GHz, see right panel of Fig.~\ref{fig:radioflux}. This decrease in the degree of polarization towards lower frequencies can be ascribed to the effect of Faraday rotation on scales smaller than angular resolution of the observations (Faraday rotation rotates the linear polarization angle proportional to the square of the wavelength, which means that the effect is stronger at lower frequencies). In general, this is known as Faraday depolarization, and provides a way to probe variations in the free electron density and magnetic field structure on scales smaller than the size of the emission region \citep{Burn1966}. 

One way to estimate the density and magnetic field variations is by using a model of external Faraday dispersion \citep{Burn1966}, described by the equation
\begin{equation}
    p = p_0 e^{-2\sigma_{\rm RM}^2\lambda^4},
\end{equation}
where $p_0$ is the intrinsic degree of polarization (i.e.~at zero wavelength), and $\sigma_{\rm RM}$ is the standard deviation of the Faraday rotation measure within the telescope beam, due to the turbulent magnetic field in the ionized gas \citep{Knuettel19} 
\begin{equation}\label{eqn:sigmaRM}
\begin{split}
\sigma_{\rm RM}&=0.83 n_e B \left( \frac{L}{{10^{13}} \rm{cm}}\, \frac{d}{10^{12} \rm{cm}} \right)^{1/2} \; \rm{rad \; m^{-2}}\\ &=256.5 \sqrt{\ln{(p_9/p_5)}} \; \rm{rad} \; \rm{m}^{-2},
\end{split}
\end{equation}
 where $n_e$ is the electron number density in cm$^{-3}$, $B$ the magnetic field strength in G, $L$ the path length through the Faraday rotating medium
 , and $d$ the scale of fluctuations. 
 The ratio of the degree of polarization between 9 and 5.5 GHz, varies from $\sim$1 (no depolarization) to $\sim$10 (strong depolarization), with an average value of $\sim 1.7$ (see Fig.~\ref{fig:radioflux}, right panel). 
 Considering the external Faraday dispersion model, a depolarization factor of 1.7 corresponds to a value of $\sigma_{\rm{RM}}\sim135$~rad~m$^{-2}$ (Eq.~\ref{eqn:sigmaRM}), which can be explained by $n_e B$ of order  $0.5$~cm$^{-3}$ G. 
 Such low values of the density and field are unlikely to be produced close to the stellar disk, which in turn suggests that the observed depolarization is produced by an extended region, where the field and density are significantly lower. 
 However, the time-variable depolarization indicates that the effect is not related to the ISM on large-scales, but instead due to an inhomogeneous medium associated with the system, possibly a clumpy, extended wind. 
We do not expect the pulsar emission to dominate the polarized signal, since depolarization in pulsars is generally not expected to be as large as observed here, due to the compact nature of the pulsar emission region \citep{sobey2019}.

\subsubsection{Pulsed radio emission from PSR B1259$-$63}
The ATCA correlator was configured such that we were able to simultaneously observe in a pulsar-binning mode at 5.5 and 9 GHz, providing an ability to measure the pulsed flux density (i.e.~from the pulsar itself), when the pulsar was visible. The initial calibration for the pulsar data followed the same procedure as the continuum. The calibrated data were then dedispersed with the {\sc miriad} task {\sc psrfix}, using a dispersion measure of 146.73 cm$^{-3}$\,pc and period \footnote{http://www.atnf.csiro.au/research/pulsar/psrcat/} of 47.7625 ms, since the ATCA data were not of sufficient time and frequency resolution to derive these values independently. 
The pulse profile of flux across the 32 bins at each frequency band was then plotted using the {\sc miriad} task {\sc psrplt}, in order to determine if the pulsed emission was visible for a particular observation (i.e.~on/off). 
{   The ATCA pulse profile is consistent with previous observations from the Parkes telescope at higher time resolution.\footnote{https://www.atnf.csiro.au/people/joh414/ppdata/1302-6350.html}
}

To obtain the pulsed flux density, we first subtracted out the continuum baseline (i.e.~emission not from the pulsar) using the {\sc miriad} task {\sc psrbl}, selecting only the bins in which the pulsar was not visible. After this, we used {\sc uvflux} on the output file to obtain the pulsed flux density and its uncertainty (in the same manner as was done for the total flux density described above). The flux density of the pulsed emission is shown in Table \ref{tab:radiodata} and Fig. \ref{fig:radioflux} (left panel) for all days for which a reliable measurement could be obtained. 
{   Initially the pulsar flux density is less than 10\% of the total flux, but beyond MJD 59300 it is a significant fraction, ranging between $\sim$30 and 50\%. We do not have measurements of the pulsar polarized flux, but it is known to be highly linearly polarized ($\sim$70 to 90\%) from the study of \cite{2002MNRAS.336.1201C}. This means that beyond MJD 59300 the pulsar polarized flux possibly contributes up to about half of the total polarized flux. We know that it cannot be all of the polarized emission also because we observed polarized flux even when the pulsar is not visible. }

\subsection{Optical Data}


Optical observations were obtained with SALT \citep{buckley06} during the 2021 periastron passage, using both the RSS and HRS instruments \citep{burgh03,bramall10}. As we noted in~\citet{Chernyakova21_psrb}, there is a hint that the equivalent width (EW) of the \halpha emission line was, on average, slightly weaker than it was during the 2014 periastron passage. For completeness we have, therefore, self-consistently remeasured the EW for the previously reported 2014, 2017 and 2021 periastron passages \citep{chernyakova15,vanSoelen16,Chernyakova20_psrb,Chernyakova21_psrb} to minimize an systematic differences that may have arisen from difference in the continuum correction or method of measurement used. A few additional points are also included post 80 days after the 2021 periastron which were not included in \citet{Chernyakova21_psrb}. All extracted spectra were continuum corrected using a low-order polynomial in the wavelength region around the \halpha emission line, and the different continuum corrections were checked against the 2021 RSS observations, for consistency. 
The EW was then measured between $\lambda = 6528 - 6607$\,\AA{} to avoid other lines in the continuum, and a linear continuum fit was used between these points to remove any residual continuum.  The EW was determined by summing over the data points (using a trapezium method) and the uncertainty was estimated following the method in \citet{vollmann06}, where the signal-to-noise ratio for each spectrum was determined following the method in \citet{stoehr08}. 

In general, the results are consistent with what has previously been reported, however, there are two changes to the 2014 observations. First, the \halpha EW for the third observation taken with the SAAO 1.9-m ($\approx 33$\,d after periaston) was found to be slightly lower. Second, was noted that for a few of the RSS observations, the peak of the emission line was slightly too bright and the CCD response was starting to become non-linear. This was realized as shorter, 20\,s exposure had been taken at the same time, and a comparison showed the core of the emission line was slightly stronger in the short exposures. However, since the first few observations did not show a difference in the line profile, this effect had not previously been noticed. 
For correctness, we checked each night and if the profile of the normalized spectrum of the short exposure was more than 2.5 per cent larger than the longer exposure (in the line core) we measured the EW from the shorter exposure.

Last, it was noted that the Modified Julian Date was recorded incorrectly by the telescope software during the 2017 observations. The time has been recalculated from the correct date and time, which shifts the time of observations by $0.5$\,d but does not change any interpretations or conclusions that were previously drawn.

\begin{figure}
    \centering
    \includegraphics[width=\columnwidth]{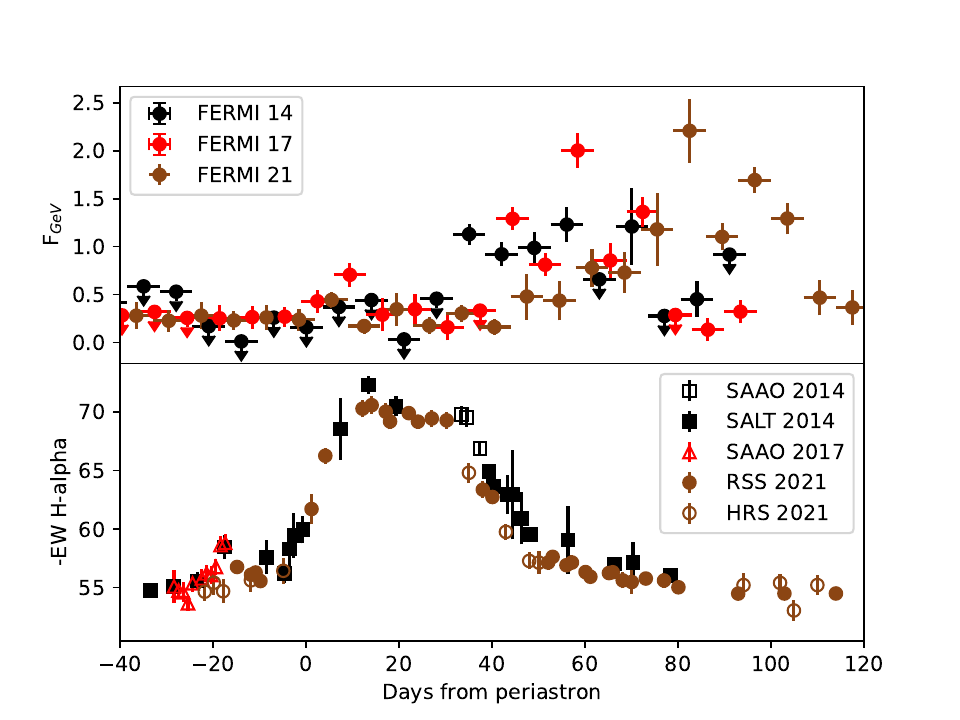}
    \caption{Evolution of GeV emission (top panel) and of the equivalent width of the \halpha emission (bottom panel) as measured over the last three periastron passages. Fermi/LAT flux measurements are given in the E $>$ 100 MeV energy range with a weekly bin size taken from \citep{Chernyakova21_psrb}. Flux is given in 10$^{-6}$ cm$^{-2}$ s$^{-1}$}.
    \label{fig:halpha_redone}
\end{figure}

The results are shown in Fig.~\ref{fig:halpha_redone} (bottom panel). This consistent re-analysis shows a remarkable similarity between the three periastron passages. The observations in 2021, show the EW of the \halpha line is slightly smaller, but the difference is not large. The increase of the line strength also starts at approximately the same time, and 
begins to decrease around 35 days after periastron.  While this time was co-incident with the start of the Fermi-LAT flare in 2014, the flare occurred much later, and no large change in line strength was observed in 2021 around the increase in GeV activity (Fig.~\ref{fig:halpha_redone}).

\subsection{X-ray Data}
 \setlength{\tabcolsep}{1pt}
\begin{table}
    \centering
    \caption{\xmm observations of \psrb. The table summarizes the Observational Id, time since periastron and best-fit parameters of the spectrum with an absorbed powerlaw model (``\texttt{cflux*tbabs*po}'') -- the total flux in 0.3-10~keV range   {(not corrected for the absorption)}, the best-fit slope and $n_H$.   {p-value shows the null hypothesis probability of the spectral fit. $\chi^2_{LC}$/ndf shows the $\chi^2$ of the fit of the lightcurve of the corresponding observation with a constant.} The moment of periastron was assumed to be $t_p=59254.867$ MJD.}

    \begin{tabular}{c|c|c|c|c|c|c}
       ObsId& $t-t_p$& Flux$_{~0.3-10}$& Index, $\Gamma$& $n_H$& $\chi^2_{LC}$/ndf & p-\\ 
         & days & $10^{-12}$erg/cm$^2$/s &  & $10^{22}$cm$^{-2}$ &  &value\\ \hline
       0881420301 &  -33 & $4.36\pm 0.05$ & $1.41\pm 0.02$ & $0.52\pm 0.01$ & 37.2/35&0.88\\ 
       0881420201  & -29  & $4.44 \pm 0.05$ & $1.38\pm 0.02$ & $0.52\pm 0.02$ & 29.9/28&0.35\\ 
       0881420401 & -25 & $3.91\pm 0.05$ & $1.32\pm 0.02$ & $0.49\pm 0.02$ & 43.8/29&0.81\\ 
       0915391301 & 753 & $0.95\pm 0.04$ & $1.71\pm 0.06$ & $0.37\pm 0.03$ & 21.3/22&0.38\\ \hline       
    \end{tabular}
    \label{tab:xmm_observations}
\end{table}
\subsubsection{\xmm}
Since the work of~\citet{Chernyakova21_psrb}, four new \xmm observations become available. We have analysed them for completeness, see Table \ref{tab:xmm_observations}.
The analysis was performed with the latest \xmm Science Analysis software v.21.0.0. Known hot pixels and electronic noise were removed, and data were filtered to exclude soft proton flares episodes   {(following the standard procedures)}. 
The spectrum was extracted from a $40''$-radius circle centred at the position of \psrb and the background was extracted from a nearby source-free region of $80''$ radius. The RMFs and ARFs were extracted using the RMFGEN and ARFGEN tools, respectively. The obtained spectra of MOS and PN cameras were fit simultaneously with the absorbed powerlaw model with the help of \texttt{XSPEC} software v.12.13.0c (\texttt{cflux*Tbabs*po} model in terms of XSPEC   {with ``wilms'' abundances~\citep{wilms_abund}}). The best-fit neutral hydrogen density $n_H$, the powerlaw photon index\footnote{Please note the difference in the definitions of radio and X-ray/GeV indexes: $\Gamma = 1-\alpha$.}  $\Gamma$ ($dN/dE\propto E^{-\Gamma}$) and $0.3-10$~keV are summarized in Table~\ref{tab:xmm_observations}.  We note that off-periastron observations at $\sim 753$~d are characterized by the softer slope and lower $n_H$ values comparing to the rest of observations taken around the first entrance to the disk. These changes could be connected to additional acceleration of the electrons as the pulsar approaches the disk of Be star and the additional absorption on the neutral hydrogen present in the disk. The obtained results are inline with previous observations reported in~\cite{Chernyakova2009}.

The   {exposure corrected} count rates detected by the PN \xmm camera are shown as a function of time in Fig.~\ref{fig:xmm_lc} for each of the considered observations. The shown light curves are binned with 600~s per point. 
None of the light curves demonstrate strong variability and are all consistent
with a constant level of flux (varying from observation to observation). The $\chi^2$ of the best-fit with a constant for each observation are summarized in the last column of Tab.~\ref{tab:xmm_observations}.

\subsubsection{Swift/XRT}
The Swift/XRT data presented in this work were adopted from~\citet{Chernyakova21_psrb}. These data include the spectral indexes in 1--10~keV energy band as well as the flux in the same energy band. In order to determine radio-to-X-ray spectral slope we explicitly assumed a power-law spectrum continuing between these energy bands. The radio-to-X-ray index was then determined based on the flux measurements in corresponding energy bands for measurements separated by no more then 1.5~days. The X-ray, radio and radio-to-X-ray spectral indexes are shown in Fig.~\ref{fig:radioflux} (middle panel) with red, green and blue points, respectively. The shown errorbars correspond to $1\sigma$ uncertainties. The Swift/XRT data similarly to \xmm were fitted with an absorbed powerlaw spectral model (\texttt{cflux*tbabs*po} in XSpec terms) with the $n_H$ absorption fixed to $0.7\times 10^{22}$~cm$^{-2}$ -- the value derived from the joint fit of all Swift/XRT observations. Please note, that the lower $n_H$ values seen in \xmm observations, see Tab.~\ref{tab:xmm_observations} would correspond to the even \textit{harder} powerlaw indexes due to the strong correlation of these parameters.

\begin{figure}
    \centering
    \includegraphics[width=\columnwidth]{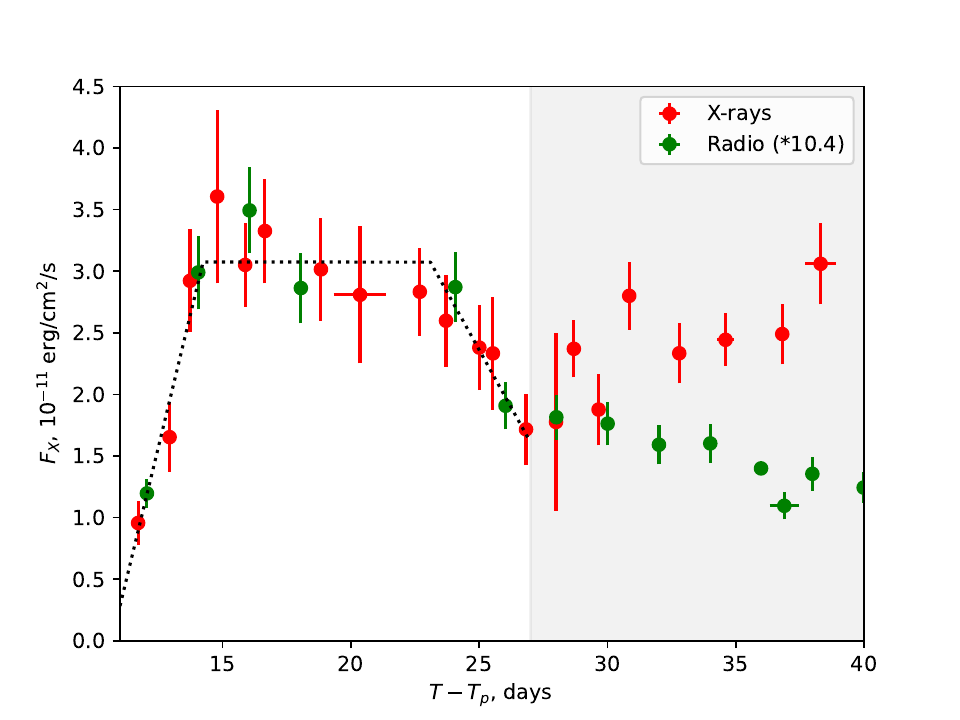}
    \caption{The X-ray (red) and radio (green) light curves of \psrb for 11 -- 40~days after the 2021 periastron passage. The light-grey shaded region shows the time when the radio/X-ray flux correlation starts to disappear. {   X-ray points are taken from  \citet{Chernyakova21_psrb}.} The dotted black curve illustrates the model used to describe the variations of the flux during the 2nd X-ray peak, see text for the details.}
    \label{fig:xray_radio_delay}
\end{figure}

\subsection{GeV Data \label{sec:GeV}}
A detailed description of our analysis of the \flat data of the \psrb during its 2021 periastron passage is presented in \citet{Chernyakova21_psrb}. 
In order to perform a proper comparison of the GeV flares during 2010, 2014, 2017 and 2021 periastron passages we reanalyzed all the data using the same version of the software (Fermitools version 2.2.0), using the latest Pass 8 reprocessed data (P8R3) from the   {SOURCE} event class   {for all event types (FRONT+BACK) with the latest IRFs (V3)}. The binned likelihood analysis was performed for photons within the energy range 0.1–100 GeV   {(summarised by $>$~0.1 GeV)} that arrived around 2010, 2014, 2017 and 2021 periastron passages within a 20$^\circ$-radius region around \psrb position,   { based on the analysis procedure recommended by  the  Fermi collaboration \footnote{https://fermi.gsfc.nasa.gov/ssc/data/analysis/} \citep[see also][]{2023hxga.book..137M}.} The selected maximum zenith angle was 90$^\circ$.   {The source model for the likelihood analysis included all sources within a 25$^\circ$-radius region around \psrb using the 4FGL-DR4 catalogue \citep{4FGL}. For the likelihood fit of the total data for any periastron, the spectral parameters of all sources outside a 5$^\circ$-radius were fixed. All other sources only had the normalization value as a free parameter, except \psrb which had a free normalization and index. For subsequent light curve and spectral points built with a likelihood fit, the index of \psrb was fixed to the previously calculated value.}

\begin{figure*}
\includegraphics[width=0.47\linewidth]{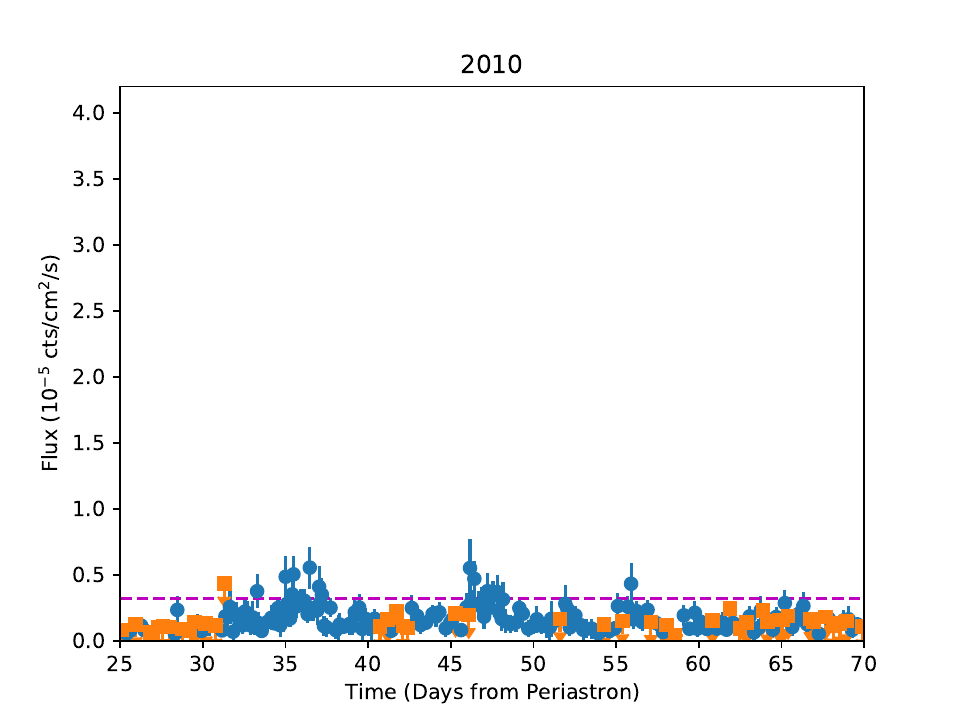}
\includegraphics[width=0.47\linewidth]{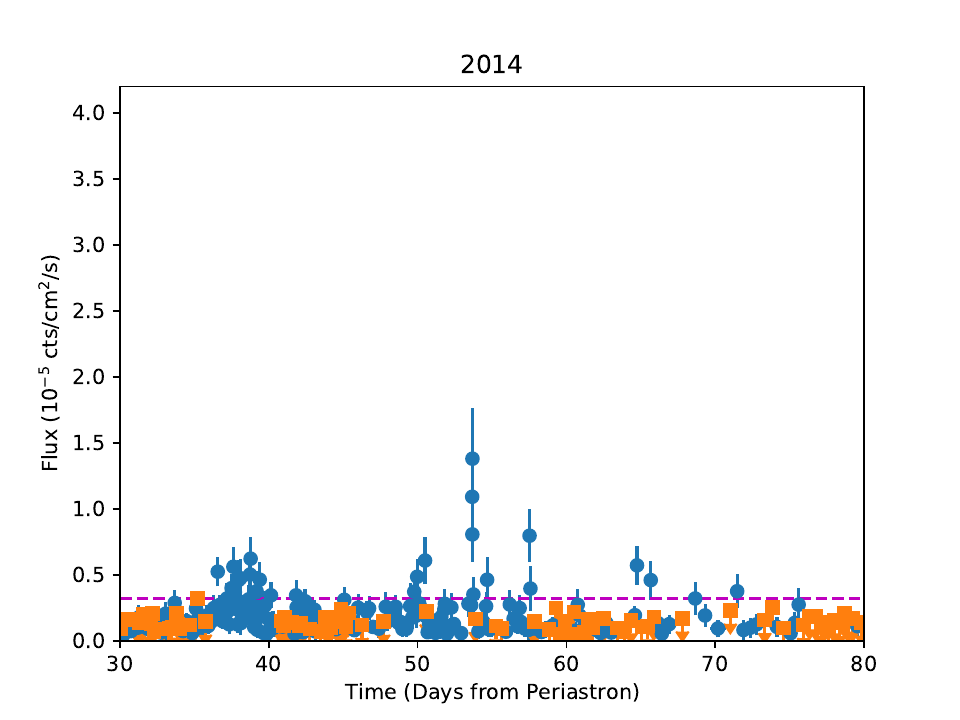}
\includegraphics[width=0.47\linewidth]{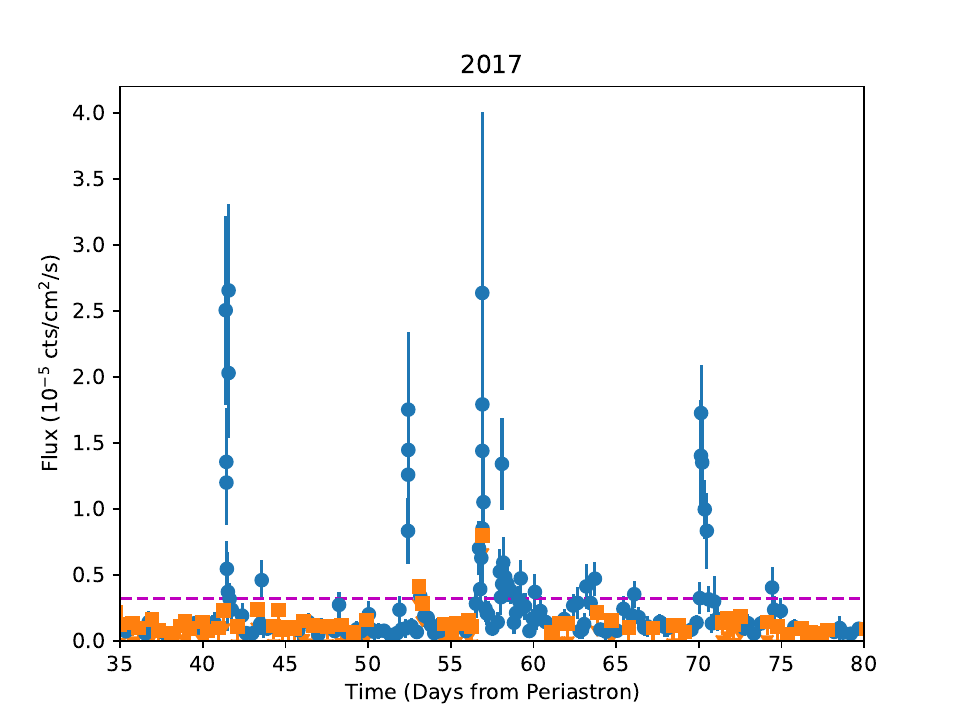}
\includegraphics[width=0.47\linewidth]{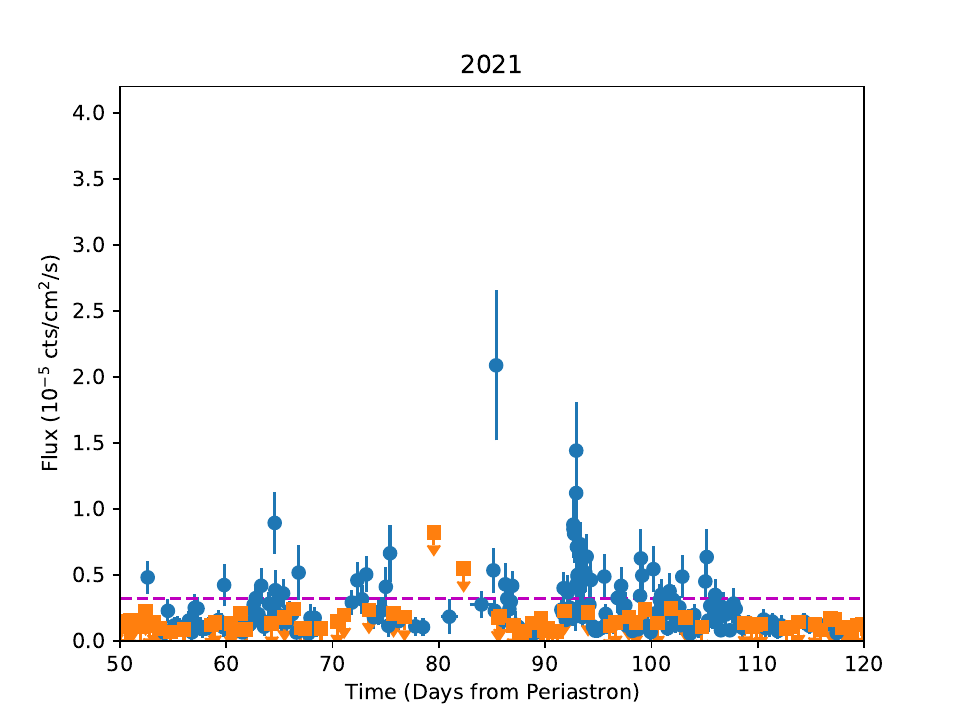}
\caption{ The GeV flares during the 2010, 2014, 2017 and 2021 periastron passages with adaptive time binning (see text). Each light curve covers the period when the highest GeV flux was detected for each periastron passage. Data points are shown in blue circles while upper limits are shown in orange squares. The horizontal dashed line indicates the isotropic flux level corresponding to the spin-down luminosity (L$_{\textrm{sd}} = 8.2 \times 10^{35}$ erg/s).} 

\label{lc1721}
\end{figure*}

  {Initially, light curves were made for each periastron period using aperture photometry. The binning was done so that each time bin accommodates 9 photons with energies above 0.1 GeV in a 1$^\circ$-radius circle around the position of \psrb. The resulting time bins have a duration of 5 min to 7 hours with an average duration of about 2 h. In Fig. \ref{lc1721} we show a comparison of the GeV light curves during the flares of different years with the same adaptive binning built through the likelihood analysis of Fermitools to verify the results of the aperture photometry. Upper limits value for both light curves and spectra were calculated with a 95\% confidence level using the IntegralUpperLimits python module available in Fermitools.} It is clearly seen from these figures that short GeV sub-flares with luminosities exceeding  the spin-down luminosity 
are present after each periastron. The number and luminosity of these sub-flares, however,  differs from periastron to periastron.

In order to compare the spectral properties of the sub-flares to the rest of the flare, we built combined spectra from all data bins with a peak flux above the spin-down luminosity and TS $>16$ (pkfl), and from all data bins below the spin-down luminosity (lowfl). These spectral parameters, for each periatron passage, are given in Table~\ref{tab:spec_all}. Also shown are the spectral parameters of the average spectrum over the flare period (avfl), and the average over the pre-flare period calculated for both 20 days before (prfl1) and 20 days after (prfl2) periastron.
For ease of comparison we fit all the data with a power-law spectrum. Note however, that at least in 2017 the prfl2 data set was better described with a cut-off power law model \citep{Chernyakova20_psrb}.

Unfortunately, the quality of the data does not allow us to unambiguously interpret the obtained results. The Tab.~\ref{tab:fermi_indexes_chi2} summarizes the results of the fit of the spectral indexes (see Tab.~\ref{tab:spec_all}) with a constant for different time periods discussed above. The periods include corresponding periastra averaged (labeled 2010--2021) and specific periods averaged over all observed periastra (labeled pfl1--avfl). The other columns indicate the best-fit $\chi^2$/d.o.f. of the fit of the spectral index with a constant; significance $\sigma$ of the consistency of the data with constant; the best-fit (average) spectral index and its uncertainty.    
Indeed, 
the significance of the variability of the average photon index  $\Gamma$, during each individual periastron passage is less than $2.5\sigma$, with the best-fit value consistent with $\Gamma=3.05$ for all periastra, except 2017. In 2017 the average slope around the periastron was significantly harder, $\Gamma_{2017}=2.74\pm0.05$.

Since the quality of the data are significantly better during the flare state, we have also tried to analyze the average spectral variability during different phases, by determining how well the slopes are fit by a constant value (Table~\ref{tab:fermi_indexes_chi2}).
We note that during the pre-flare (prfl1 and prfl2), and low-flare (lowfl) periods, the spectral indices are consistent with a constant value.  However, the peak-flare (pkfl) and average flare (avfl) data sets are not consistent with a constant spectral index at a $\sim3\sigma$ and a $\sim 4\sigma$ level, respectively (Tab.~\ref{tab:fermi_indexes_chi2}).  One can also conclude that, on average, the spectrum of the source during the period of GeV flare is softer than before the flare.

\begin{table}

    \centering
    \caption{Spectral parameters around 2010, 2014, 2017 and 2021 periastron passages when fitted with a power-law model ($dN/dE\propto E^{-\Gamma}$). The second column gives either the dates of the observations relative to periastron or the total length of the state if it is shown in brackets. The last column gives values of test statistics of the source detection during the given period. For the definition of different periods see the text. }
    \begin{tabular}{c|c|c|c|c|}
     \hline
	Data set & Period (d)& $\Gamma$   & Flux ($>$0.1 GeV)& TS\\
             &(Duration, d)&        &   $10^{-6}$ ph cm$^{-2}$ s$^{-1}$                &\\
      \hline

2010-prfl1&-20 -- 0 &	3.47$\pm$ 0.58 & 0.34$\pm$ 0.08& 	29.2\\
2014-prfl1&-20 -- 0 &	2.70& 	$<$0.25& 	4.2\\
2017-prfl1&-20 -- 0 &	2.82$\pm$ 0.29& 0.27 	$\pm$ 0.07 & 	23.2\\
2021-prfl1&-20 -- 0 &	2.49$\pm$ 0.40& 	0.15$\pm$ 0.10&	13.4\\
\hline
2010-prfl2&0 -- 20 &	2.91$\pm$ 0.18& 0.38$\pm$ 0.06& 	53.5\\
2014-prfl2& 0 -- 20&	2.52$\pm$ 0.22 & 0.16$\pm$ 	0.06& 	15.6\\
2017-prfl2& 0 -- 20&	2.78$\pm$ 	0.13& 	0.49$\pm$ 0.07& 77.5\\
2021-prfl2& 0 -- 20&	2.84$\pm$  	0.52& 	0.30$\pm$  	0.08& 	39.7\\
\hline
2010-lowfl& (39.1) &	3.00$\pm$0.07& 	1.27$\pm$ 	0.07& 	725.6\\
2014-lowfl& (39.4) &	3.22$\pm$0.09&	1.05$\pm$0.06& 	692.4\\
2017-lowfl& (31.1) &	2.87$\pm$ 0.09& 0.95$\pm$ 	0.07& 	367.5\\
2021-lowfl &  (48.1)	&3.12$\pm$ 0.07& 1.34$\pm$ 	0.06& 	754.3\\
\hline
2010-pkfl& (0.4) &2.92$\pm$ 	0.29& 	4.64$\pm$ 	0.63& 	110.6\\
2010-avfl&30 - 70 &	3.04$\pm$	0.07& 	1.29$\pm$ 	0.06& 	796.6\\
2014-pkfl& (1.4)&3.08$\pm$ 	0.10& 	5.12$\pm$	0.32& 	512.1\\
2014-avfl& 35 - 76&	3.16$\pm$ 	0.07& 	1.17$\pm$ 0.05& 	907.1\\
2017-pkfl& (1.8)&	2.64$\pm$ 	0.07& 	7.43$\pm$ 	0.39& 	785.6\\
2017-avfl&41 - 75 &	2.79$\pm$ 	0.05& 	1.28$\pm$	0.07& 	729.6\\
2021-pkfl& (3.2) &2.86$\pm$ 	0.08& 	5.23$\pm$ 	0.27 &	740.2\\
2021-avfl&55 - 108 &	3.00$\pm$ 	0.05& 	1.56$\pm$ 	0.06& 	1161.1\\
\hline

    \end{tabular} 
    
    \label{tab:spec_all}
\end{table}

\begin{table}

    \centering
    \caption{ The $\chi^2$ value for each period if the Fermi indexes are fit with a constant value.  The significance of the data not being described by a constant, is given by $\sigma$.} 
    \begin{tabular}{cccc}
     \hline
     Period & $\chi^2$/d.o.f & $\sigma$& Best-fit $\Gamma$ \\ \hline
     2010 & \quad0.96/3\quad~ & \quad1.3\quad~ & $2.99 \pm 0.06$ \\
     2014 & 8.77/2 & 2.5 & $3.10 \pm 0.06$\\
     2017 & 4.30/3 & 1.2 & $2.74 \pm 0.05$ \\
     2021 & 7.72/3 & 1.9 & $3.00 \pm 0.05$ \\ \hline
     prfl1 & 1.94/2 & 0.5 &$2.81 \pm 0.22$\\
     prfl2 & 1.92/3 & 0.8 &$2.77 \pm 0.09$\\
     lowfl & 9.07/3 & 2.2 & $3.05 \pm 0.04$\\
     pkfl & 13.72/3 & 3.0 &  $2.81 \pm 0.05$\\
     avfl & 20.95/3 & 4.0 & $2.95 \pm 0.03$\\ \hline
     \end{tabular}
     \label{tab:fermi_indexes_chi2}
\end{table}

\begin{figure}
\includegraphics[width=0.9\linewidth]{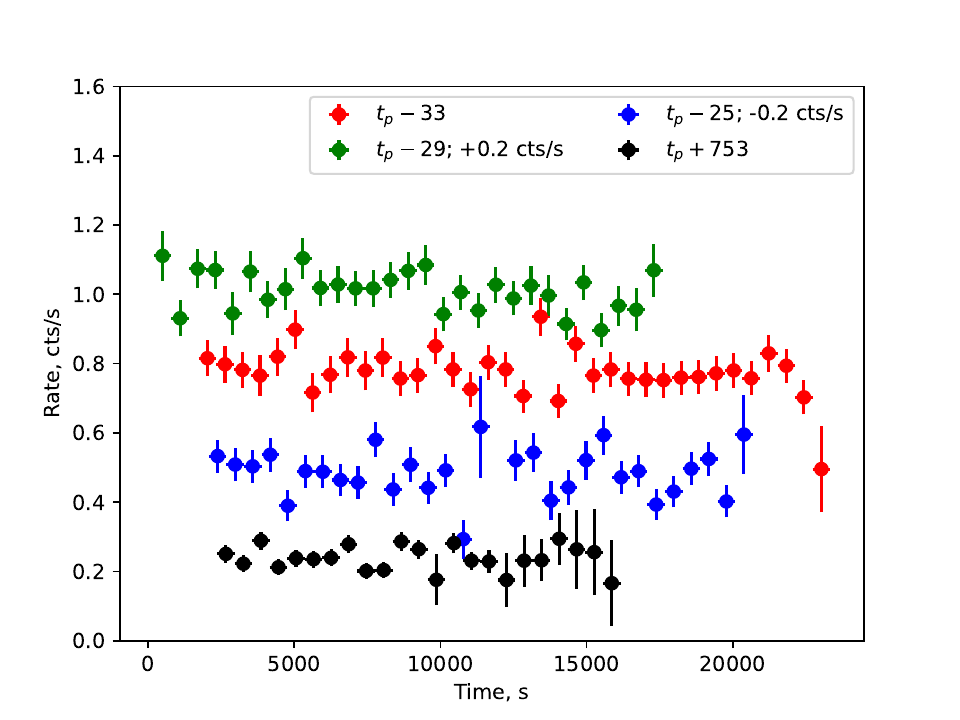}
\caption{The X-ray flux variability during the observations performed by \xmm, see Table~\ref{tab:xmm_observations}. The duration of each point is 10 minutes. All light curves show the count rate detected by the PN camera.}
\label{fig:xmm_lc}
\end{figure}

\section{Modelling}
\label{sec:model}

In our analysis we have used the same model 
as was used to describe the data from 2017 periastra passages in \citet{Chernyakova20_psrb}. In this model we assume that the observed emission is coming from two main populations of relativistic electrons: \textit{(i)} electrons of the unshocked and weakly shocked pulsar wind (producing GeV emission via IC and/or bremsstrahlung mechanisms)  and \textit{(ii)} strongly shocked electrons accelerated near the apex of the shock (producing X-ray and TeV emission via synchrotron/IC mechanisms), see Fig.~5 of~\citet{Chernyakova20_psrb}.  In this model one can expect to see 
an enhancement of the GeV emission when the bow shock cone is pointing 
in the direction of the observer, see Appendix~\ref{sec:appendix} for more details.
  {Please note that in the discussed 2-zone toy model several potentially important effects (anisotropy of the IC emission; exact geometry of the system) are not explicitly considered. The model is not designed to calculate the exact variability of the emission along the orbit, but rather to illustrate that the observed enhancement of the emission can be understood in general. In our simplified model we have calculated all the IC emission on the soft photons of the Be star only ($T=27\,000$ K), and assumed the same photon density in both emission zones.}

 The spectrum of unshocked electrons was selected to be a power law with a slope of $\Gamma_e=-2$ in the energy range $E_e=0.2-1$~GeV.  A small fraction of electrons are additionally accelerated at the strong shock near the apex  to $E_e\sim 500$~TeV energies with the similar slope $\Gamma_e=-2$. The rest of the electrons flying into the shock direction are reverted to flow along the shock at the surface of stellar-pulsar wind interaction cone far from the apex, and could be additionally mildly accelerated on a weak shock. This leads to a power law tail in the spectrum of diverted electrons with a softer slope $\Gamma_2$ which continues above $1$~GeV to at least $E_e\sim 5$~GeV. A slope which is softer than 2 is a characteristic of particle acceleration on weak shocks~\citep[see e.g.][]{bell78,blandford87}. 

 After the injection, the spectra of both populations are  modified by radiative (Inverse Compton, synchrotron, or bremsstrahlung) and non-radiative (adiabatic or escape) losses operating in the system. In our calculations, the resulting electron spectrum was determined by numerically
calculating the radiative losses of a continuously injected spectrum
of electrons, until a steady solution has been obtained. The time that electrons spend in the emitting region is about a few thousand seconds, and is long enough to substantially modify the injected spectrum.

\begin{figure*}
\includegraphics[width=0.45\linewidth]{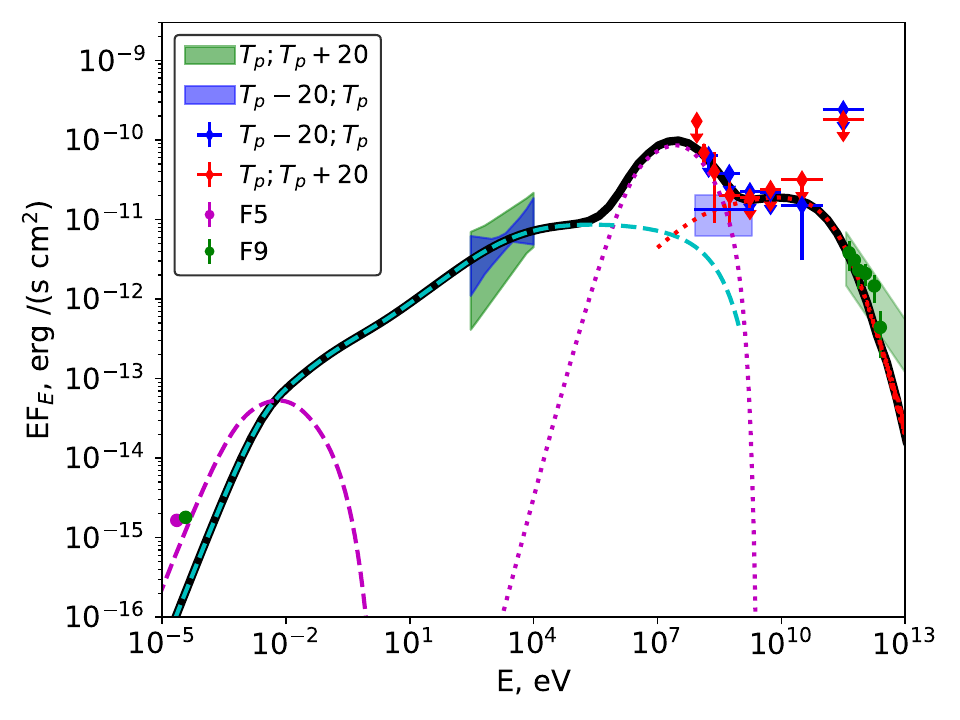}
\includegraphics[width=0.45\linewidth]{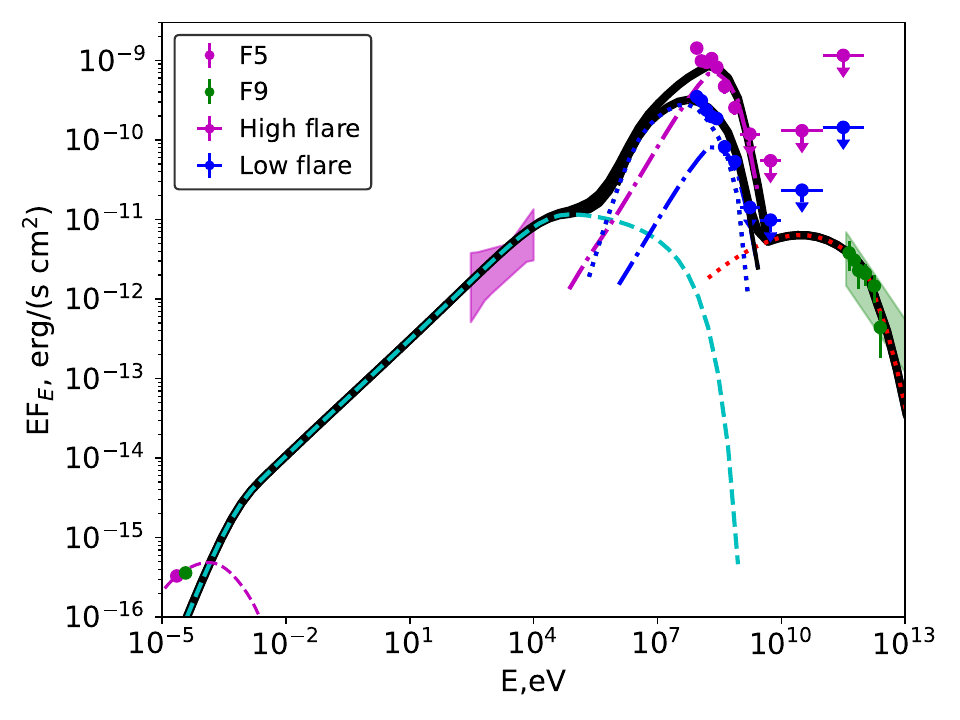}
\caption{The GeV emission of \psrb during (\textit{left:} )  twenty days before (blue points, prfl1) and after (red points, prfl2) the periastron, and (\textit{right:}) during the 2021 flaring period. Magenta points correspond to the high state of the flare (2021-pkfl) and blue points to the low state during the flare (2021-lowfl). Dashed lines represent the contribution of synchrotron emission, dotted lines the IC emission, and dash-dotted lines the bremsstrahlung emission.   {Note that the contribution of the IC component in the GeV region is the same for both the low and high states of the flare (dotted blue curve).} The radio points F5 and F9  in the left panel correspond to radio emission at 5 and 9.5 GHz 20 days after the periastron, and 50 days after periastron in the right panel. The blue, green and magenta butterflies in the keV range correspond to the range of the observed X-ray emission by SWIFT/XRT 20 days before and after the periastron, and during the period of GeV flare, respectively. 
The TeV points are taken from \citep{2005A&A...442....1A}, and the shaded regions at TeV energies represent the range of multiyear H.E.S.S.  measurements \citep[see Fig. 2 of][]{hess_psrb2020}.  }
\label{fig:sed}
\end{figure*}

In the analysis of the broadband data of 2021 periastron passage we considered separately the spectra during the 20 days before (prfl1) and  after (prfl2) the periastron passage, and during the high (pkfl) and  low (lowfl) flaring states. We found that the model  describes the data reasonably well, and the contribution of bremsstrahlung emission is important only during the flare period, see Table~\ref{tab_mods} and Fig.~\ref{fig:sed}. We note that for the modelling of the flaring state we have assumed the escape time to be rather short, $t_{esc}=1000$\,s in order to match to the duration of the shortest subflares discussed in section \ref{sec:GeV}.  
We would like to note also a strong correlation between modeled clumps' density $n_{clump}$ and the escape time from the system $t_{esc}$. In order to maintain the total energy in the relativistic electrons present in the system to be roughly the same, the longer $t_{esc}$ requires lower values of $n_{clump}$.

  {Note that all results discussed above are strongly model dependent and are used here as general illustration of the model. Moreover, the parameters of the model (magnetic filed strength, soft photon field density, electrons spectrum, escape time) are strongly correlated and form a degenerate parameter space. Thus the observations can be explained via somewhat different model parameters.  More accurate modelling taking into account the full geometry of the system is needed for a  better reconstruction of system parameters.}



 \begin{table}
\caption{Details of the model parameters, see section \ref{sec:model}. D is a distance from the Be star to the emission region, $\Gamma_2$ is a slope of accelerated electrons above 1 GeV, and $t_{esc}$ is time the electrons stay in the emitting region before the escape. Effective luminosity L of the pulsar wind electrons (without considering  beaming effects) is measured in units of spin-down luminosity L$_{sd}=8.2\times10^{35}$ erg/s.}
\label{tab_mods}
\begin{center}
\begin{small}
	\begin{tabular}{|c|c|c|c|c|c|c|}
 \hline
	Data & D   &   n$_{clump}$   &B  &$\Gamma_2$ & t$_\textrm{esc}$ & L/ L$_{sd}$         \\ 
	Period &10$^{13}$cm& 10$^{10}$cm$^{-3}$& G& &s& \\\hline
	pkfl&5.5&40&0.2&5&1000&30\\\hline
	lowfl &5.5&0.4&0.2&4&4000&30\\\hline
 prfl2& 1.5&$\lesssim$ 0.1&1.& 3&6000&1\\
 \hline
    
\end{tabular}
\end{small}
\end{center}

\end{table}
 
\section{Radio to X-ray behavior}
The relation between the radio and X-ray components in \psrb has a non-trivial behavior. As it was reported in \citet{Chernyakova21_psrb} at first the radio observations corresponding to the time of the second X-ray peak (from $\sim$ 11 to 27 days after the 2021 periastron passage) show a clear correlation with the X-ray flux. After this time the radio to X-ray correlation stops, see Fig.~\ref{fig:xray_radio_delay}.
Using the data obtained during the correlation period we performed a search for a possible delay between the X-ray and radio emission (expected, for example, in the case when the emission in these bands is produced in spatially separated regions). For this purpose we selected the time interval of 11-27~days after the periastron passage (white region in Fig.~\ref{fig:xray_radio_delay}) and re-normalized the radio-band light curve to match the average X-ray flux during this time period. We modelled both the X-ray and re-normalized radio light curves with a linear increase -- constant -- linear decrease of the flux model, see dotted black curve in Fig.~\ref{fig:xray_radio_delay}. In addition, we suggested that exactly the same model describes both X-ray and radio data sets except that the model describing the radio data is delayed in time in relation to the X-ray data. We fit this model to both the X-ray and radio data sets simultaneously, assuming the flux increase/decrease rates, the start/end times of the constant flux level and the radio-X-ray flux delay to be free parameters. The best-fit of the model to the data is shown with dotted black line in Fig.~\ref{fig:xray_radio_delay}. We did not find any significant delay between X-ray and radio emissions with a best-fit delay value of $0\pm 0.20$~d.

In terms of the spectral-energy distribution, the radio data in general lie on the continuation of the X-ray spectrum (Figures \ref{fig:radioflux} \& \ref{fig:sed}), though the spectral index in the radio band is systematically softer than the index in the X-ray band and the index that would be required to simultaneously fit the radio and X-ray data.

The fact that observations do not show a single power-law from radio to X-rays is not surprising. Within our proposed model the energetics constraints do not allow the population of electrons to produce a single, continuous power-law spectrum of photons from the radio to the X-ray band.

  {
The radio emission could be partially explained as synchrotron emission from the same population of electrons which produces the GeV emission via IC/bremsstrahlung mechanisms. However, the predicted spectral slope in this case significantly differs from the observed one that does not allow us to explain all the observed radio emission in this way.  The dashed magenta curve in the left panel of Fig.~\ref{fig:sed} illustrates the emission in a characteristic magnetic field of $\sim 0.2$~G around the periastron. Note that any higher value of the magnetic field in the radio emission region will produce emission at a higher level than the observed one and is thus disfavored by our model.  During the GeV flare a two orders of magnitude lower magnetic field is required. Note that the exact value is strongly model dependent and can be higher if one takes into account possible absorption of the radio emission or/and assume higher clump density and attributing most of the GeV emission to bremsstrahlung emission.  }


We argue that the majority of observed radio emission is produced either by the high-energy electrons of the pulsar wind that have cooled down during their movement along the shock front, or by a different population of electrons, e.g. electrons in the Be-star's wind which are accelerated at the shock.

In the case that the radio emission is produced by the cooled down population of the relativistic electrons, the characteristic properties of the population and the magnetic field  can be estimated as follow. The characteristic frequency of the photons emitted by electrons with an energy $E_e$ in a magnetic field $B$ is:
\begin{align}
& \nu_{\rm synch}\simeq 10 \left(\frac{B}{0.1\,\mbox{G}} \right) \left(\frac{E_e}{100 \, {\rm MeV}}\right)^2 \mbox{GHz.}    
\label{eq:synch_freq}
\end{align}
The corresponding cooling time for the same electrons is:
\begin{align}
& t_{\rm synch}= \frac{6 \pi m_e c}{\sigma_T B^2 \gamma}\simeq 10 \left(\frac{B}{0.1\,\mbox{G}}\right)^{-2} \left( \frac{E_e}{100\,\mbox{ MeV}}\right)^{-1} \mbox{yr.}\label{eq:tcool}
\end{align}
 The available radio-band observations indicate variability on time scales as short as $\sim 1$~day. If the radio emission is produced by the cooled down population of relativistic electrons, this variability time scale should not exceed the cooling time, i.e. $t_{\rm synch}\lesssim 1$~d. Together with this, the requirement that  $\nu_{\rm synch}$ is equal to the radio frequency of the observations ($\nu_{\rm synch}\sim 10$~GHz) provides two equations for the two unknowns ($E_e$ and $B$). The solution given by the equations results in unreasonably high values of the magnetic field $B\sim24$\,G.    { Such high values of the magnetic field would indicate that the emission region is located much closer to the star (to keep the ratio of the energy densities of the magnetic and photon fields at the same level), but will not strongly affect the spectral shape of the electron population.
 The requirement of the high magnetic field could be overcome in the case when the escape time of the electrons is  significantly shorter than $t_{synch}$. In this case, however, the cooling of the electrons is very inefficient and it is
difficult to explain that the full energy release in the radio band is roughly consistent with the continuation of the X-ray spectrum. }

The introduction of a separate population of electrons does not require high magnetic fields, but instead requires a fine tuning to explain the similarities of the X-ray and radio spectral indexes as well as the similarities of the radio flux and the power-law continuation of the X-ray spectrum. The origin of the initial correlation between radio and X-ray emission and its further disappearance \citep{Chernyakova21_psrb} is also not clear in this case.


\section{Discussion and conclusions}
In the current paper we present the results of the analysis of the radio-to-GeV data for the recent periastron passage of 2021 and systematic re-analysis of all previous GeV (\flat) and optical (SAAO/SALT) data. 

We present here for the first time the behaviour of \psrb at 9 GHz, the radio spectral index, and degree of polarization at both 5.5 and 9 GHz during the 2021 periastron passage (Table \ref{tab:radiodata}). 
The total radio flux is initially well correlated with the X-ray emission (i.e.~the second X-ray peak), however the third X-ray peak is uncorrelated with the radio emission, which continues decreasing at both 5 and 9 GHz. 
The spectral index of the radio emission is consistent with optically thin synchrotron emission, with no evidence for free-free absorption during the observed period. In general, the radio spectrum is inconsistent with a single population of electrons explaining both X-ray and radio emission. 
The degree of polarization of the radio emission increases substantially (from $\lesssim$5 to $\sim$25\%) as the pulsar emerges from behind the disk, as it experiences less depolarization overall and/or the magnetic field in the emission region becomes more ordered. Time-variable depolarization is also observed, which is most likely due to a clumpy, extended wind associated with the system but not close to the stellar disk (where the field and density are too high to explain the observations). 

The consistent re-analysis shows a remarkable similarity between the behaviour of the EW of the \halpha lines during last three periastron passages.   {Many Be X-ray binary systems show superorbital periods, linked to variations in the circumstellar disc \citep[e.g.][]{rajoelimanana11}, and this is found for the gamma-ray binary LS\,I +61\,303 \citep[e.g.][]{paredes15}. The lack of large changes in the equivalent width seems to suggest a similar outflow from the star, and a consistent behaviour for the disc over the last few periastron passages. } The observed differences in optical ranges are much smaller than the difference in the GeV behavior. While the decrease in line strength was co-incident with the start of the \flat flare in 2014, the flare in 2021 occurred much later, with no large change in line strength around the start of GeV activity, see Fig.~\ref{fig:halpha_redone}.  Similar to the previous report of  \citet{Chernyakova21_psrb} we do not find any counterparts of the 2021 GeV flare at any other wavelengths. 
We note, however, that the binary separation at the time of the GeV flare is $\approx3 - 5$\,AU. Such distances significantly exceed the size of the region where the majority of the \halpha emission is expected to arise. This suggests that the GeV flare could manifest 
  {itself at longer wavelengths \citep[e.g.][]{klement17}}.
Such observations can be more sensitive to the changes in the colder, outer regions of the Be star disk and would be of extreme importance for the identification of GeV flare counterpart at other wavelengths. 

In this paper we also present analysis of all recent (after 2021) \xmm observations, see Table \ref{tab:xmm_observations} and Fig. \ref{fig:xmm_lc}. We note that, similar to previous periastron passages, the spectral slope of the system is significantly harder shortly before the periastron ($\Gamma \sim 1.3 - 1.4$) than near the apastron  ($\Gamma\sim 1.7$). We do not see any significant variability on short (10 minutes) time scale, see Fig.~\ref{fig:xmm_lc}.  At the same time we found that the spectral index in the radio band is systematically softer than the one in the X-ray band, or the index derived from the interpolation between radio and X-ray energies. We also do not find any evidence for a delay between X-ray and radio emission with the best-fit delay to be $0\pm 0.20$~d.

A systematic comparison of timing and spectral properties of  GeV flares over the last 4 periastron passages demonstrates that all flares differ in their total duration,  overall short-timescale structure and spectral properties, see Tables \ref{tab:spec_all} and \ref{tab:fermi_indexes_chi2}.

Our model describes the X-ray-to-TeV data reasonably well with parameters similar to presented before \citep{Chernyakova20_psrb}, see Table \ref{tab_mods}. At the same time this model  under predicts the radio flux (see Figure \ref{fig:sed}). The radio data seems to be located on the powerlaw continuation of the X-ray data, although the X-ray-to-radio index is systematically harder 
than radio one, see Figure \ref{fig:radioflux}. To explain this one may consider either effective cooling (requires very high magnetic field B$\sim 20$\,G) or a second population of electrons (e.g. electrons of the Be-star's wind accelerated at the shock). The introduction of a second electron population would need a fine tuning to explain the initial  radio/X-ray correlation and its subsequent disappearance  as well as the fact that the observed radio emission is lying roughly on the continuation of the X-ray spectrum.  The assumption of a very high magnetic field would mean that the magnetic field in the emitting region is, at least at some orbital phases, dominated by the magnetic field of the Be star \citep[as assumed in e.g.][]{Melatos95}, which can lead to interesting consequences, e.g. correlation of X-ray/TeV emission (as both magnetic and soft photon fields are proportional to the distance from the star).

\section*{Acknowledgements}
 DM is supported by DFG through the grant MA 7807/2-1 and DLR through the grant 50OR2104. MC and SMK acknowledge support of ESA Prodex grant C4000120711. SPO acknowledges support from the Comunidad de Madrid Atracción de Talento program via grant 2022-T1/TIC-23797. The authors acknowledge support by the state of Baden-W\"urttemberg through~bwHPC. The research conducted in this publication was jointly funded by the Irish Research Council under the IRC Ulysses Scheme 2021 and ministères français de l’Europe et des affaires étrangères (MEAE) et de l'enseignement supérieur et de la recherche (MESR). BvS acknowledges support by the National Research Foundation of South Africa. This paper uses observations made at the South African Astronomical Observatory (SAAO). Some of the observations reported in this paper were obtained with the Southern African Large Telescope (SALT) under programmes 2018-1-MLT-002 (PI: B. van Soelen) and 2021-2-LSP-001 (PI: DAH Buckley).

\section*{Data Availability}
The data underlying this article will be shared on reasonable request to the corresponding authors.

\input{journals.tex}
\bibliography{references.bib} 

\appendix

\section{Enhancement of the GeV emission}
\label{sec:appendix}

The only way to explain the fact that during the GeV flare we observe sub-flares with the luminosity greatly exceeding the spin down one is to take into account the anisotropy of the emission. 

Below we show the back of the envelope estimations allowing us to estimate the flux magnification factor $\alpha$ as a function of opening angle of the emitting cone $\Omega_{ani}$, the characteristic angular size of the clump $\Omega_{cl}$ and the Lorentz factor $\gamma$  of the relativistic electrons crossing the clump. We show that the flux magnification of $\sim 30$ seen for the brightest \flat subflares could be produced by clumps with the characteristic angular scale (as seen from the apex of the cone) is $\sim 10^{-2}$~sr.


In the case that the source is not emitting isotropically into 4$\pi$, but rather into a solid angle $\Omega_{ani}$, then the number of electrons is linked to the isotropic case as 
\[
N_{real,el}=N_{iso,el} \times 4\pi/\Omega_{ani}
\]

The number of electrons intercepted by a clump with the angular size $\Omega_{cl}$ (as seen from the cone's apex) is:

\[
N_{cl,el}=N_{real,el}  \frac {\Omega_{cl}}{ \Omega_{ani}} = N_{iso,el} \times 4\pi   \frac {\Omega_{cl}} {\Omega_{ani}^2}
\]



The interaction of the electrons with the clump will lead to the Bremsstrahlung emission. 
The number of photons produced via this emission is proportional to the number of electrons crossing the clump:
\[
N_{cl,ph}=\kappa N_{cl,el}
\]

Note that the photons produced in this interaction
are not emitted isotropically, but confined into a cone with an opening angle

\[
\Omega_{cl,ph}=max\{\Omega_{cl}, \Omega_{emi}\} 
\]
Here $\Omega_{emi}$ corresponds to the opening angle of emission cone due to Doppler boosting. The apex angle of such emission is $\theta_{emi}\sim 1/\gamma$. 
The corresponding solid angle is
\[
\Omega_{emi} = 4 \pi \sin^2 \frac {\theta_{emi}} {2} = \frac{\pi}{\gamma^2}
\]



The number of photons that reach the observer from the clump is 
\begin{align}
& N_{obs,ph} = \frac{\Omega_{obs}}{\Omega_{cl,ph}} N_{cl,ph} \\ \nonumber
& \Omega_{obs} = \frac{A}{D^2} \\ \nonumber
& F_{cl,ph} = N_{obs,ph}/A 
\end{align}


Here $\Omega_{obs}$ is the angular size of the observer (as seen from the clump), $A$ -- the effective area of the observer's instrument, $D$ -- the distance between the clump and the observer.

Substituting all the equations above one could express the flux seen by the observer $F_{obs,ph} = N_{obs, ph}/A$ as a function of the flux expected for the case of isotropic photon emission in the system $F_{iso,ph}=N_{iso, ph}/(4\pi D^2)$:
\begin{align}
&    F_{obs,ph} = \frac{N_{obs,ph}}{A} = \frac{N_{cl,ph}}{\Omega_{cl,ph}D^2}  = \\ \nonumber
& = \frac{\kappa N_{cl, el}}{\Omega_{cl,ph}D^2} = \frac{4\pi\kappa N_{iso, el} \Omega_{cl}}{\Omega^2_{ani} \Omega_{cl,ph}D^2} = \\ \nonumber
&=\frac{N_{ph,iso}}{4\pi D^2}\cdot\frac{16\pi^2\Omega_{cl}}{\Omega^2_{ani} \Omega_{cl,ph}} = F_{iso, ph}\cdot \frac{16\pi^2\Omega_{cl}}{\Omega^2_{ani} \Omega_{cl,ph}} 
\end{align}


The bulk gamma-factor of the relativistic electrons typically seen in simulations is $\gamma\sim 2-3$ \citep[e.g.][]{Bogovalov19}, while the clumps are usually suggested to be small and located relatively far from the cone's apex. In this case $\Omega_{emi} > \Omega_{cl}$ and $\Omega_{cl,ph}\approx \Omega_{emi}$. Then


\begin{align}
& F_{obs,ph} = F_{iso,ph}\cdot \frac{16\pi^2\Omega_{cl}}{\Omega^2_{ani} \Omega_{cl,ph}}\approx   F_{iso,ph}\cdot \frac{16\pi^2\Omega_{cl}}{\Omega^2_{ani} \Omega_{emi}} = \\ \nonumber
& = F_{iso,ph}\cdot 16\pi\gamma^2\frac{\Omega_{cl}}{\Omega^2_{ani}}
\end{align}

The flux magnification factor is then
\begin{align}
& \alpha\equiv\frac{F_{obs,ph}}{F_{iso,ph}} = 16\pi\gamma^2\frac{\Omega_{cl}}{\Omega^2_{ani}}    
\end{align}


Now we can estimate the size of the clump needed to produce the bright flares observed to greatly overcome the spin-down luminosity. 
$\Omega_{ani}$ can be estimated from the length of the time-period when happens the bright subflares, $\theta_{ani} \sim 20^\circ$, hence $\Omega_{ani}\sim 0.4$. Then


\[
\Omega_{cl} = \frac{1}{30\pi} \left(\frac{\Omega_{ani}}{0.4 }\right)^2 \frac{\alpha}{30} \frac{9}{\gamma^2}
\]
i.e. $\Omega_{cl} \sim 1/30\pi$ for the typical values of the assumed parameters ($\alpha\sim 30$, $\gamma\sim 3$). We note also that the discussed condition $\Omega_{emi}>\Omega_{cl}$ indeed holds as $\Omega_{emi}\approx \pi/\gamma^2 > 1/30\pi = \Omega_{cl}$.




We note also, that in case if the clump penetrates through both shock waves and interacts with the electrons of the unshocked wind, then the flux of electrons intersected by the clump from the isotropic central source would be
\[
F_{cl,el}=F_{iso,el} \frac{\Omega_{cl}}{4 \pi}
\]

The observed photon flux from the clump (under the assumption that in this case $\Omega_{cl} > \Omega_{emi}$) is equal to
\[
F_{obs}=\frac{N_{obs}}{A}=\frac{N_{cl, ph}}{A}\frac{\Omega_{obs}}{\Omega_{cl}}= \frac{N_{cl, ph}}{D^2 \Omega_{cl}}
\]

\[
F_{obs}=\frac{\kappa N_{cl,el}}{D^2 \Omega_{cl}}=\frac{\kappa N_{iso,el} \Omega_{cl}}{4\pi D^2 \Omega_{cl}}=\kappa F_{iso,el}=F_{iso,ph}
\]

This means that in this case the flux of the photons detected by the observer cannot exceed the spin-down luminosity.

\end{document}

%% file: journals.tex
\def\aj{AJ}%
\def\actaa{Acta Astron.}%
\def\araa{ARA\&A}%
\def\apj{ApJ}%
\def\apjl{ApJ}%
\def\apjs{ApJS}%
\def\ao{Appl.~Opt.}%
\def\apss{Ap\&SS}%
\def\aap{A\&A}%
\def\aapr{A\&A~Rev.}%
\def\aaps{A\&AS}%
\def\azh{AZh}%
\def\baas{BAAS}%
\def\bac{Bull. astr. Inst. Czechosl.}%
\def\caa{Chinese Astron. Astrophys.}%
\def\cjaa{Chinese J. Astron. Astrophys.}%
\def\icarus{Icarus}%
\def\jcap{J. Cosmology Astropart. Phys.}%
\def\jrasc{JRASC}%
\def\mnras{MNRAS}%
\def\memras{MmRAS}%
\def\na{New A}%
\def\nar{New A Rev.}%
\def\pasa{PASA}%
\def\pra{Phys.~Rev.~A}%
\def\prb{Phys.~Rev.~B}%
\def\prc{Phys.~Rev.~C}%
\def\prd{Phys.~Rev.~D}%
\def\pre{Phys.~Rev.~E}%
\def\prl{Phys.~Rev.~Lett.}%
\def\pasp{PASP}%
\def\pasj{PASJ}%
\def\qjras{QJRAS}%
\def\rmxaa{Rev. Mexicana Astron. Astrofis.}%
\def\skytel{S\&T}%
\def\solphys{Sol.~Phys.}%
\def\sovast{Soviet~Ast.}%
\def\ssr{Space~Sci.~Rev.}%
\def\zap{ZAp}%
\def\nat{Nature}%
\def\iaucirc{IAU~Circ.}%
\def\aplett{Astrophys.~Lett.}%
\def\apspr{Astrophys.~Space~Phys.~Res.}%
\def\bain{Bull.~Astron.~Inst.~Netherlands}%
\def\fcp{Fund.~Cosmic~Phys.}%
\def\gca{Geochim.~Cosmochim.~Acta}%
\def\grl{Geophys.~Res.~Lett.}%
\def\jcp{J.~Chem.~Phys.}%
\def\jgr{J.~Geophys.~Res.}%
\def\jqsrt{J.~Quant.~Spec.~Radiat.~Transf.}%
\def\memsai{Mem.~Soc.~Astron.~Italiana}%
\def\nphysa{Nucl.~Phys.~A}%
\def\physrep{Phys.~Rep.}%
\def\physscr{Phys.~Scr}%
\def\planss{Planet.~Space~Sci.}%
\def\procspie{Proc.~SPIE}%
\let\astap=\aap
\let\apjlett=\apjl
\let\apjsupp=\apjs
\let\applopt=\ao